\documentclass[12pt]{article}

\usepackage{epsfig}

\begin{document}

\title{Regge and factorized GPD models in $\rho^0$ elastic electroproduction}

\author{A.V. Vinnikov\footnote{E-mail: vinnikov@theor.jinr.ru} \\
{\small \it 141980, BLTP, Joint Institute for Nuclear Research,
Dubna, Russia }}

\date{\today}

\maketitle

\begin{abstract}A possibility to study Regge and factorized
parameterizations of generalized parton distributions
in hard electroproduction of $\rho^0$ on the proton
is considered. For that the dependences of the differential cross
sections on the transferred momentum for these parameterizations
are compared.
\end{abstract}

In the past few years, much attention was drawn to generalized parton
distributions (GPDs) \cite{gpds,markus}. This is related to
important issues in hadron physics which the knowledge of GPDs can
resolve: nucleon spin composition \cite{jirule} and three-dimensional
nucleon structure \cite{impact}. GPDs formalism provides as well
a consequent description of dynamics of hard {\it exclusive} hadronic
reactions in terms of perturbative QCD.

At present, several experiments planned to study
hard exclusive electroproductions of photons and mesons are expected to
give access to GPDs \cite{h1,zeus,hermes,clas}.
However, the extraction of information on GPDs from the
electroproduction data is not simple. For example, the amplitude
of photon electroproduction involves not only the deeply virtual Compton
scattering process (DVCS) described in terms of GPDs, but also
a contribution (dominant in a wide range of kinematics) from
the Bethe-Heitler process. Thus, taking into account experimental
uncertainties, it is not easy to extract pure DVCS contribution to
the observables. In meson production, the meson wave function
must be taken into account, which again gives rise to the
uncertainty. Moreover, it was shown \cite{vdh} that higher twist
effects in hard exclusive meson electroproduction can be very large.
Taking into account transverse motion of quarks in the proton
and in the meson yields a factor of 5 suppression to the cross
section at $Q^2=4$ GeV$^2$ and a factor of 2 suppression at
$Q^2=10$ GeV$^2$.
Therefore, estimation of model uncertainties in description
of hard exclusive electroproduction reactions is important for
extraction of GPDs from experimental data.
In this letter, an impact of the input GPDs model on the cross section
of $\rho^0$ electroproduction is considered. In particular, results
of the cross section calculation in Regge and factorized GPDs
models are compared. This issue is raised since Regge behavior of the
GPDs is now widely discussed \cite{markusreg,vandreg,lat1,lat2,lat3}

The cross section of hard exclusive $\rho^0$ electroproduction reads
\cite{vdh}:
\begin{equation}
\frac{d \sigma_L}{dt}=\frac{|{\mathcal T}|^2}
{8m\pi(W^2-m^2)|{\vec q}_{1}| }
\end{equation}
where ${\vec q}_1$ is the momentum of the photon in the center of mass
system of the photon and the initial proton, $W$ is their
invariant mass, $m$ is the proton mass, $t$ is the transferred momentum and
${\mathcal T}$ is the scattering amplitude.
The perturbative QCD describes only production of the longitudinal
vector mesons by the longitudinal photons \cite{frankstrik97}
therefore the present calculations cover only the longitudinal
photons and $\rho^0$ mesons. In this letter, the most
simple description for the amplitude is chosen in order to check
the impact on the calculated cross section from the input GPD model solely.
Therefore, the transverse motion effects are omitted even though
they are expected to be large. In this approximation,
the amplitude reads\footnote{Here, only the spin non-flip amplitude
is taken into account. The spin-flip one calculated within the model
of Ref.~\cite{gpv} gives very small contribution to the cross section
and becomes visible only when considereing asymmetries.}:
\begin{equation}
{\mathcal T}=-\imath e \frac{2\sqrt{2}\pi\alpha_s}
{9 Q}{\mathcal A} {\bar u}(p_2)n^{\alpha}\gamma_{\alpha}u(p_1)
\int\limits_0^1 dz \frac{\Phi(z)}{z}.
\end{equation}
Here $Q=\sqrt{-q_1^2}$, $n=(1,0,0,-1)/(\sqrt{2}(p_1+p_2)^+)$ is a light-like
vector along the $z$-axis, $p_1$ and $p_2$ are 4-momenta
of the initial and final protons correspondingly.
The $\rho^0$-meson wave function is taken in the form
\begin{equation}
\Phi(z)=6z(1-z)f_{\rho}
\end{equation}
with $f_{\rho}$=0.216 GeV$^2$. The factor ${\mathcal A}$ is given by
\begin{equation}
{\mathcal A}=\frac{1}{\sqrt{2}}\int\limits_{-1}^{1}\Bigl ( e_u H_u(x,\xi,t)
-e_d H_d(x,\xi,t) -
\frac{3}{8}(e_u-e_d)\frac{H_g(x,\xi,t)}{x} \Bigr )
\left \{ \frac{1}{x-\xi+\imath\epsilon}+\frac{1}{x+\xi-\imath\epsilon}
\right \} dx,
\end{equation}
where $\xi=Q^2/4(p_1\cdot q_1)$.
The functions $H(x,\xi,t)$ entering the above expressions are
the spin non-flip GPDs\footnote{In this leter, GPDs definitions
of \cite{markus} are used.} of quarks ($u,~d$) and gluons ($g$).

At present GPDs are usually described by simple parameterizations
providing general symmetry properties (e.g. polynomiality) and
reducing to the conventional parton
distribution functions (PDFs) in the forward limit.
The most common form for these parameterizations
is the double distribution \cite{radyushkin} complemented with the
D-term \cite{dterm1}. This scheme allows to relate
the non-forward distributions to the conventional (forward) PDFs
and the proton elastic form factors. In this framework,
the quark GPD $H$ is given by
\begin{equation}
H_q(x,\xi,t)=\frac{4m^2-2.8 t}{4m^2-t}\frac{H_q(x,\xi)}
{(1-t/0.71)^2}.
\end{equation}
The $t$-independent part of the GPD reads
\begin{equation}
H_q(x, \xi)\,=\,H_q^{DD}(x, \xi) \,+\,
\theta(\xi-|x|)\, \frac{1}{N_f} \, D(\frac{x}{\xi})\, ,
\end{equation}
where $H_q^{DD}$ is the part of the GPD which is obtained
from the double distribution (DD) $F_q$:
\begin{equation}
H_q^{DD}(x,\xi)=
\int\limits_{-1}^{1}d\beta\
\int\limits_{-1+|\beta|}^{1-|\beta|} d\alpha\
\delta(x-\beta-\alpha\xi)\  F_q(\beta,\alpha)\ .
\end{equation}
For the double distributions Radyushkin's suggestion \cite{radyushkin}
is used
\begin{equation}
F_q(\beta, \alpha) = \frac{\Gamma(2b+2)}{2^{2b+1}\Gamma^2(b+1)}\
\frac{\bigl[(1-|\beta|)^2-\alpha^2\bigr]^{b}}{(1-|\beta|)^{2b+1}}q(\beta).
\label{quarkdd}
\end{equation}
At $x>0$, $q(\beta)=q_{val}+\bar q $ is the forward quark density for the
flavor $q$.
The negative $x$ range corresponds to the antiquark density:
$q(-x)=-{\bar q}(x)$. For the profile parameter $b$ the value 1.0 is taken.
For the PDFs, the MRST98 \cite{mrst98} parameterization at $Q^2$=4 GeV$^2$
was used.

The $D$-term can be presented as series of Gegenbauer polynomials.
The first terms were calculated in Ref. \cite{gpv}:
\begin{equation}
D(z)=(1-z^2)\biggl[ -4.0 \ C_1^{3/2}(z)
-1.2 \ C_3^{3/2}(z) -0.4\ C_5^{3/2}(z) \biggr]\ .
\end{equation}

For the gluons, the spin non-flip distribution is taken as
\begin{equation}
H_g^{DD}(x,\xi)=
\int\limits_{-1}^{1}d\beta
\int\limits_{-1+|\beta|}^{1-|\beta|} d\alpha
\delta(x-\beta-\alpha\xi)  \beta F_g(\beta,\alpha),
\end{equation}
with the same shape of the profile functions in the double distribution
\begin{equation}
F_g(\beta, \alpha) = \frac{\Gamma(2b+2)}{2^{2b+1}\Gamma^2(b+1)}
\frac{\bigl[(1-|\beta|)^2-\alpha^2\bigr]^{b}}{(1-|\beta|)^{2b+1}}
g(\beta) ,
\label{gluondd}
\end{equation}
and $b$=2. The $t$-dependence for the gluons is taken the same
as for the quarks.

Another type of GPD parameterization is the Regge ansatz, where
the double distributions (Eqs.\ref{quarkdd},\ref{gluondd})
have a Regge-type form:
\begin{equation}
F_{q,g}(\beta,\alpha,t)=F_{q,g}(\beta,\alpha)\frac{1}{|\beta |^{\alpha' t}}.
\end{equation}
Here $\alpha '$ is the slope of the Regge trajectory. For quarks,
$\alpha ' = 1$ GeV$^{-2}$, for gluons $\alpha ' = 0.25$
GeV$^{-2}$. Note that this parameterization is very simple
and it is used in the present calculation since its main goal is
to estimate, rather then to calculate carefully the effect of
Regge $t$-dependence. For solid analysis of Regge ansatz in GPD modelling
see Ref.~\cite{markusreg,vandreg}.

\begin{figure}[htb]
\centering
\begin{minipage}[c]{0.35\hsize}
\epsfig{file=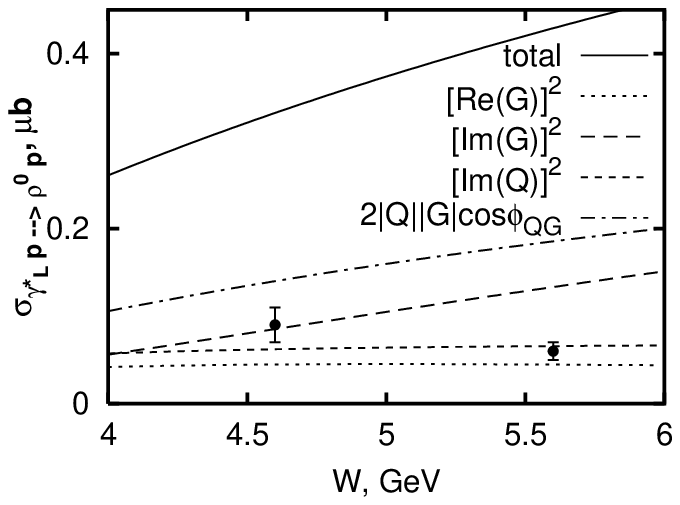,width=\hsize}
\end{minipage}
\begin{minipage}[c]{0.35\hsize}
\epsfig{file=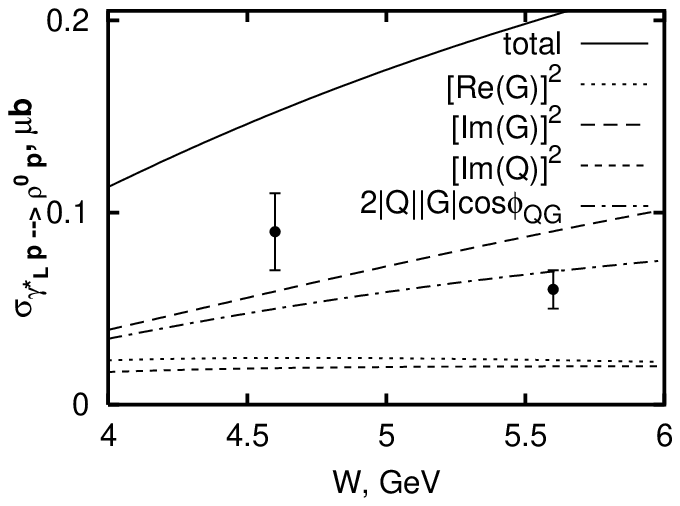,width=\hsize}
\end{minipage}
\caption{The calculated hard exclusive $\rho^0$ electroproduction
cross section at $Q^2=4 $ GeV$^2$ for the factorized (left)
and Regge (right) GPDs models versus HERMES \cite{rhoherm} data.
$Q$ is quark amplitude, $G$ is gluon amplitude and $2|Q||G|\cos\phi_{QG}$
is their interference. The quark real part is very small and is not shown.}
\label{rhocross}
\end{figure}

The dependence of the cross section on the energy for
the photon virtuality $Q^2$=4 GeV$^2$ is shown in Fig.\ref{rhocross}.
The calculated cross section overshoots considerably the experimental
data. This could be expected since the parton transverse motion
was not taken into account. Besides, a significant impact on the value
of the cross section can arise from the choice of the scale
in $\alpha_s$ \cite{blm}.
For both factorized and Regge input GPDs the present result
confirms the conclusion of Ref.~\cite{markusandi} that the ratio of
gluonic and quark contributions to $\rho^0$ electroproduction
is large even at intermediate energies. 

\begin{figure}[htb]
\centering
\begin{minipage}[c]{0.35\hsize}
\epsfig{file=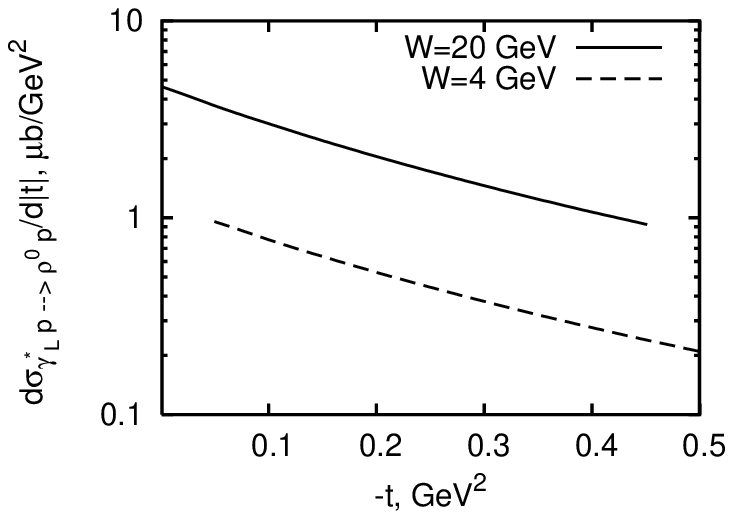,width=\hsize}
\end{minipage}
\begin{minipage}[c]{0.35\hsize}
\epsfig{file=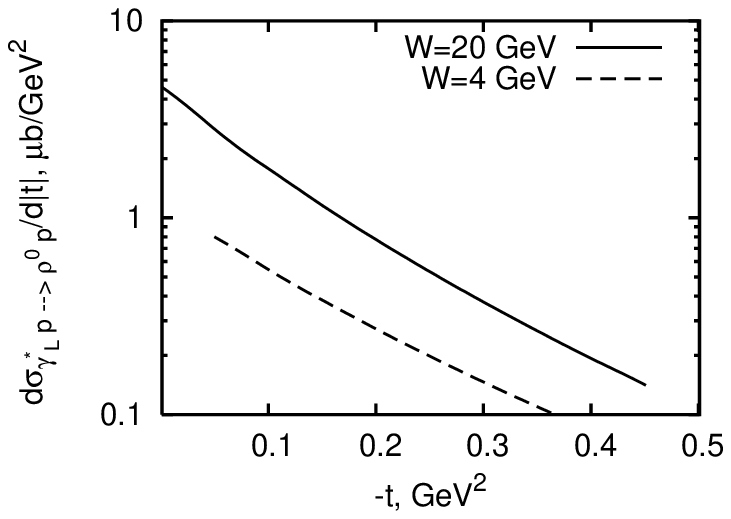,width=\hsize}
\end{minipage}
\caption{The differential cross section of hard exclusive
$\rho^0$ electroproduction at $Q^2=4 $ GeV$^2$ for the factorized (left)
and Regge (right) GPDs models.}
\label{tdep}
\end{figure}                                                                   

From Fig.~\ref{rhocross} it is clear that Regge
and factorized models of the GPD $H$ lead
to visibly different values of the cross section.
The reason of it is the well-known Regge exponential
suppression of the differential
cross section (see Fig.~\ref{tdep}). However, such a different
behaviors of the cross sections does not allow for experimental
distinction of the veritable model (or extraction the value
of the slope $\alpha '$ from data). Indeed,
the momentum transfer goes mostly in the transverse direction:
\begin{equation}
t=-4\xi^2m^2-(p_2-p_1)_{\perp}^2\approx - (p_2-p_1)_{\perp}^2.
\end{equation}
The larger the transverse momentum transfer is, the stronger is
suppression due to the intrinsic transverse motion.
Therefore it is expected that even in the factorized model a significant
differential cross section suppression should occur as the
momentum transfer grows. It means that including
effects of the intrinsic transverse motion
or inserting Regge behavior in the GPD $t$-dependence
give a similar effect for $\rho^0$ electroproduction and a way
to disentangle these two has to be found to access GPDs.

It is well-known that in soft reactions Regge behavior of the
differential cross sections is expressed not only in the exponential
dependence on $t$ but also in the shrinkage. This means that
as the c.m. energy grows, the $t$-dependence of the
differential cross section becomes steeper. 
Thus the shrinkage
might become an appropriate effect to investigate the
$t$-dependence of the GPDs. However, a numerical calculation
shows that the shrinkage does not help to study the $t$-dependence
of the GPD $H$ in $\rho^0$ electroproduction.
Indeed, the $t$-dependence for two
different energies ($W$ = 4 and 20 GeV) given in Fig.~\ref{tdep}
shows almost no shrinkage. This is related to the different
slopes $\alpha '$ for quarks and gluons. At larger energy, the gluons
are becoming dominant and the ``effective'' Regge slope of the
GPD $H$ becomes smaller. Therefore, the effect of shrinkage is depreciated.

To conclude, it appears that study of Regge behavior of the GPD $H$
in hard exclusive $\rho^0$ electroproduction is embarrass by the
effects of intrinsic transverse motion of partons in the nucleon
and in the meson. A study of the intrinsic transverse motion
and a calculation of its impact on the
differential cross section is necessary to access the $t$-dependence
of the GPD $H$ in this reaction.

The author is gratefull to W.-D.~Nowak, M.~Diehl, Z.~Ye and F.~Ellinghaus
for discussions. This work is supported by RFBR grants 03-02-17291
and 04-02-16445.

\end{document}